\documentclass[%
 reprint,
 amsmath,amssymb,
 aps,
 pre,
]{revtex4-2}

\usepackage{graphicx}% Include figure files
\usepackage{dcolumn}% Align table columns on decimal point
\usepackage{bm}% bold math
\usepackage{lipsum}
\usepackage{color}

\def\<{{\langle}}
\def\>{{\rangle}}
\def\g2{{\gamma_2}}
\def\pt{\partial_t}
\def\pg{\partial_g}
\def\pgg{\partial^2_{gg}}
\def\f2{\frac{1}{2}}
\def\pmu{\partial_{\mu}}
\def\pnu{\partial_{\nu}}
\def\pmn{\partial^2_{\mu \nu}}

\def\bg{\bm{g}}
\def\ve{\varepsilon}

\begin{document}

\preprint{APS/123-QED}

\title{Dynamic landscapes and statistical limits on growth during cell fate specification}% Force line breaks with \\

\author{Gautam Reddy}
 \email{greddy@princeton.edu}
\affiliation{%
 Joseph Henry Laboratories of Physics, Princeton University,
Princeton, New Jersey 08544, USA\\
}%


\begin{abstract}
The complexity of gene regulatory networks in multicellular organisms makes interpretable low-dimensional models highly desirable. An attractive geometric picture, attributed to Waddington, visualizes the differentiation of a cell into diverse functional types as gradient flow on a dynamic potential landscape. However, it is unclear under what biological constraints this metaphor is mathematically precise. Here, we show that growth-maximizing regulatory strategies that guide a single cell to a target distribution of cell types are described by time-dependent potential landscapes under certain generic growth-control tradeoffs. Our analysis leads to a sharp bound on the time it takes for a population to grow to a target distribution of a certain size. We show how the framework can be used to compute regulatory strategies and growth curves in an illustrative model of growth and differentiation. The theory suggests a conceptual link between nonequilibrium thermodynamics and cellular decision-making during development. 
%Our analysis highlights the interplay between gene regulation, stochastic gene expression and tradeoffs that impact cellular growth during development. 
\end{abstract}

\maketitle

%\tableofcontents

% TODO:

% \begin{enumerate}
%     \item Read and add more references from development.
%     \item Expand on the $h\Sigma = -\alpha I$ assumption. 
%     \item Clarify the Sinkhorn algorithm
%     \item Clarify the thermodynamic connection
% \end{enumerate}

Organismal development is a tremendously complex process involving the coordinated growth and differentiation of a single cell into a well-defined population of cell types. During cell fate specification, the re-organization of gene expression profiles is coordinated by complex regulatory mechanisms that parse external signals and control the expression of hundreds to thousands of genes \cite{davidson2010regulatory, levine2005gene, reik2007stability}. Both contextual instructive signals and stochastic factors influence the eventual fates of a cell and its descendants \cite{wernet2006stochastic, losick2008stochasticity, symmons2016s}.  

%The conflicting demands of rapid growth and differentiation during development or healing thus impose constraints on how quickly an organism matures or re-establishes homeostasis. 

%Cellular growth and differentiation are coordinated by complex regulatory networks that control the expression of hundreds to thousands of genes. The eventual fates of a cell and its descendants are influenced both by external instructive signals and intrinsic randomness. 
%A striking example is the diversification of R7 and R8 photoreceptor cells during \emph{Drosophila} retinal development into two distinct subtypes. The specification of the two types is determined by the stochastic expression of a single regulatory gene in R7 cells, whose dynamics produces a defined ratio of the two subtypes in the population. The stochastic fate of the R7 cell in turn instructs the fate of its partner R8 cell.
One would hope that cellular processes involved in cell fate specification can be described by interpretable models that reflect core regulatory principles and guide new experiments. One such intuitive picture, provided by Waddington, is that of a ball (an undifferentiated cell) rolling down a dynamic potential landscape with a ``valley'' bifurcating into multiple valleys, corresponding to the distinct fates that the cell can acquire \cite{waddington2014strategy, ferrell2012bistability}. Under what physiological and functional constraints is this metaphor a mathematically precise description of cellular decision-making? One perspective emphasizes the structural stability property of gradient-like dynamical systems to motivate Waddington-like low-dimensional models of cell fate decisions \cite{saez2022dynamical,saez2022statistically,rand2021geometry,raju2023geometrical, freedman2023dynamical}. A related static picture views cell types as attractors in energy-based models of associative memory \cite{lang2014epigenetic,yampolskaya2023sctop,boukacem2024waddington}.  

Another perspective is provided by optimal transport \cite{villani2009optimal, santambrogio2015optimal}, which offers a powerful computational framework for tracing single-cell gene expression profiles over time \cite{schiebinger2019optimal, bunne2023learning, bunne2024optimal}. Conceptually, optimal transport considers the context-dependent transformation of an initial distribution of cell states to a final distribution of cell states. Different optimal transport formulations correspond to different assumptions on the biological cost of transforming cell states. 

The celebrated Benamou-Brenier theorem \cite{brenier1991polar, benamou2000computational} bridges these two perspectives by showing that under certain quadratic transport costs, the optimal transport map is described by gradient flow on a time-dependent potential landscape. %The Benamou-Brenier formalism is also closely related to thermodynamic speed limits in stochastic thermodynamics \cite{aurell2011optimal,aurell2012refined,van2023thermodynamic} and path integral formulations of stochastic control \cite{chen2016relation, chen2021stochastic, kappen2005path, todorov2009efficient}. 
However, it is unclear how quadratic (or other) costs on transport maps relate to physiologically relevant constraints, bringing into question the interpretation of maps inferred by optimal transport. Moreover, growth is central to development. Existing frameworks either ignore growth or relax the hard requirement that mass is conserved to a soft constraint when analyzing data \cite{chizat2018unbalanced, schiebinger2019optimal}.   

\begin{figure}[t!]
    \center
    \includegraphics[scale=0.4]{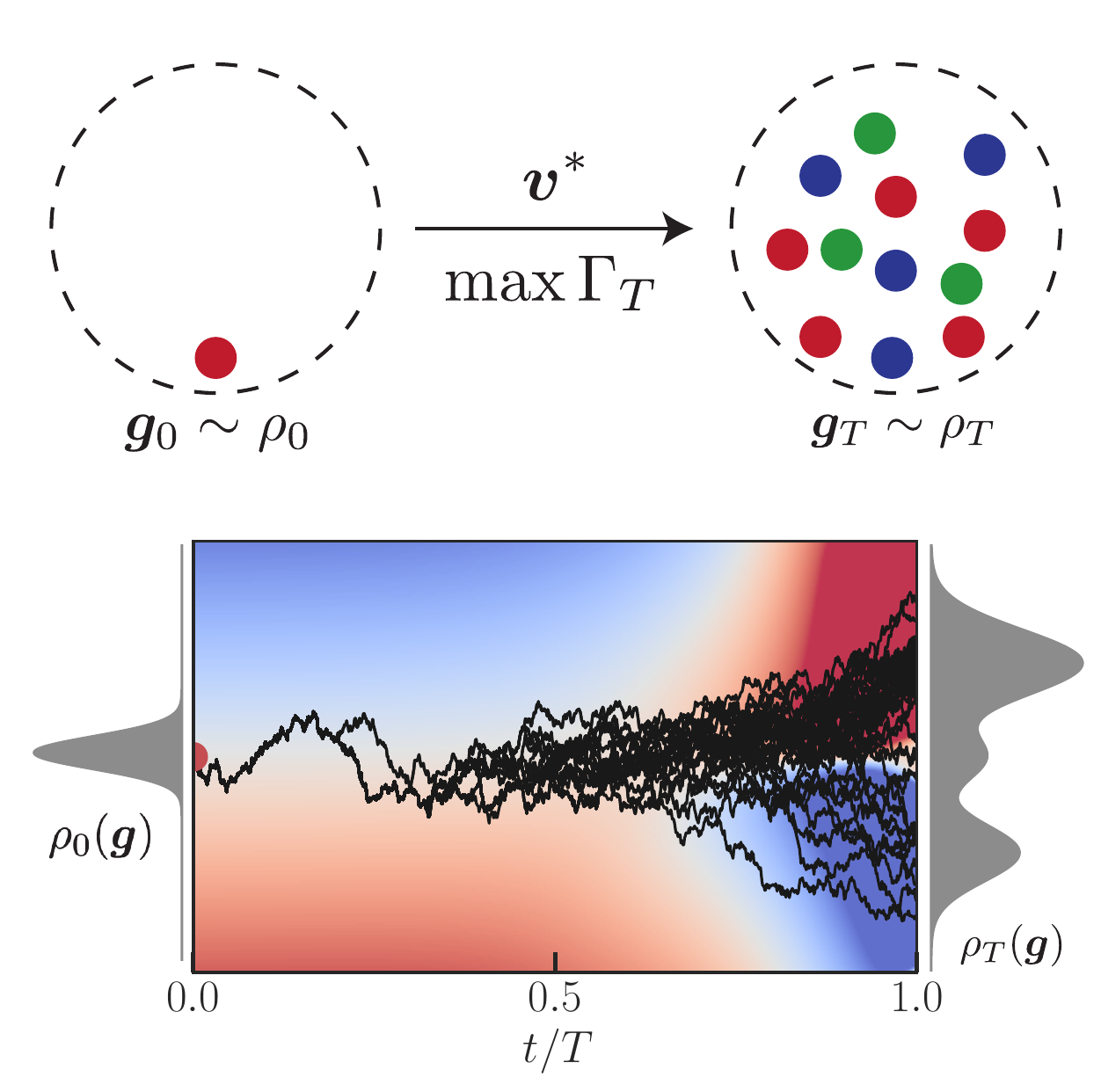}
    \caption{We consider the growth of a cell with a multi-dimensional cellular state $\bg$ from an arbitrary probability density $\rho_0(\bg)$ to a population with arbitrary target density $\rho_T(\bg)$ in finite time $T$. The optimal regulatory control $\bm{v}^*$ maximizes average growth rate $\Gamma_T = T^{-1} \log N_T/N_0$, where $N_T, N_0$ are the final and initial numbers of cells, respectively. Below, a single cell replicates and its descendants acquire distinct fates due to stochasticity in cellular state dynamics. The optimal regulatory control $v^*(g,t)$ is shown in red ($v > 0$) and blue ($v < 0$).}
    \label{fig:fig1}
\end{figure}

Here, we consider a population of non-interacting cells growing into an arbitrary target distribution of cell states in a finite time while maximizing growth. An example is presented in Figure \ref{fig:fig1}. We present four main contributions: (1) An analog of the Benamou-Brenier result for a certain class of growth-control tradeoffs, which identifies mathematical constraints under which a Waddington-like picture is valid (Section \ref{sec:maintheorem}). (2) An expression for the density of the optimally controlled process, which motivates numerical methods for computing the potential landscape (Section \ref{sec:maintheorem}). (3) An exact expression for the potential landscape when growth begins from a single cell with a known state (Section \ref{sec:single}). To illustrate the modeling framework, we use this exact expression to derive the optimal regulatory strategy that guides the proliferation and differentiation of a fast-growing cell type into a mixture of slow and fast growing cell types on a one-dimensional epigenetic landscape (Section \ref{sec:1d_main}). (4) A precise statistical bound on population growth rate analogous to bounds on entropy production (``speed limits'') in nonequilibrium thermodynamics (Section \ref{sec:bound}).

The framework further motivates the development of computational transport methods that incorporate physiologically relevant constraints. A notable departure from existing approaches is that gene regulation has a direct impact on growth, and all transport costs are built into this tradeoff. The framework is generally applicable to scenarios where diversifying into a heterogeneous population conflicts with maximizing instantaneous growth. Examples include microbial bet-hedging in unpredictable environments \cite{ackermann2015functional, veening2008bistability}, and the rapid proliferation and differentiation of cells into specialized types during inflammation and wound healing \cite{landen2016transition}.

\section{Model}
We assume each cell has a state $\bm{g}$ (for example, its gene expression profile) which evolves as
\begin{align}
    dg_{\mu} = \left(f_{\mu}(\bm{g},t) + v_{\mu}(\bm{g},t)\right)dt + \sigma_{\mu \nu} dW_{\nu},\label{eq:gmu}
\end{align}
where $\mu$ indexes the components of $\bm{g}$, $f_{\mu}(\bm{g},t)$ describes the passive dynamics of $g_{\mu}$ (for example, dilution and degradation), $v_{\mu}(\bg,t)$ is the regulatory control variable and $dW_{\mu}$ is a Wiener process that represents noise due to stochastic gene expression and other factors (note $\langle dW_{\mu}(t) \rangle = 0$, $\langle dW_{\mu}(t) dW_{\nu}(t')\rangle  = \delta_{\mu \nu} \delta(t - t') dt$ and Einstein summation is used throughout). Here, time $t$ represents the time since an external triggering event, but in general reflects the influence of a time-varying external signal on gene regulation $\bm{v}$.  

The cell doubles with a probability per unit time (i.e., growth rate) $\gamma(\bg,\bm{v})$. This joint dependence on $\bg$ and $\bm{v}$ reflects the fact that a cell's growth rate, for example, may depend on how much of the proteome is devoted to protein synthesis and the fraction of the synthesis machinery that is occupied by the synthesis of proteins that do not contribute towards further synthesis. We consider terms to second-order in $\bm{v}$:
\begin{align}
    \gamma(\bm{g},\bm{v}) = \gamma_0(\bm{g}) + d_{\mu}(\bm{g})v_{\mu} + \f2 h_{\mu \nu}(\bm{g})v_{\mu} v_{\nu}, \label{eq:gamma}
\end{align}
where $\gamma_0(\bg)$ and $d_{\mu}(\bg)$ are arbitrary functions of $\bg$ and $h_{\mu \nu}$ is symmetric with additional constraints discussed further below. $d_{\mu}$ is set to zero by noting that it can be recovered by an appropriate translation of $f_{\mu}, v_{\mu}$ and $\gamma_0$. A cell's progeny when it doubles inherit the same state as their parent but their states will subsequently diverge due to noise (Figure 1). From \eqref{eq:gmu} and \eqref{eq:gamma}, the (unnormalized) density $\rho(\bg, t)$ satisfies the forward equation
\begin{align}
    \pt \rho  + \pmu \left((f_{\mu} + v_{\mu}) \rho \right) - \f2 \Sigma_{\mu \nu} \pmn \rho - \gamma \rho = 0. \label{eq:rho}
\end{align}

We aim to find the growth-maximizing regulatory strategy $\bm{v}^*(\bg,t)$ that guides a population of non-interacting cells with an arbitrary (but known) initial probability density $\rho_0$ to an arbitrary target probability density $\rho_T$ in finite time $T$. 

\section{Background}
Our modeling framework is closely related to the optimal transport framework \cite{villani2009optimal, chewi2024statistical, peyre2019computational}, and the generalized Schrödinger bridge problem \cite{schrodinger1931umkehrung, schrodinger1932theorie, chen2021stochastic, chetrite2021schrodinger, gentil2017analogy}. Optimal transport is concerned with finding the joint density $\Pi(\bg_0,\bg_T)$ with the constraint that the marginal distributions of $\Pi$ are $\rho_0$ and $\rho_T$ respectively. The problem is ill-posed without an additional objective that specifies the cost $c(\bg_0,\bg_T)$ of moving probability mass from an initial state $\bm{g}_0$ to the final state $\bm{g}_T$. That is, optimal transport finds $\Pi$ that minimizes $\int d\bg_0 d\bg_T c(\bg_0,\bg_T) \Pi(\bg_0,\bg_T)$ subject to the constraints $\int d\bg_0  \Pi(\bg_0,\bg_T) = \rho_T(\bg_T)$ and $\int d\bg_T  \Pi(\bg_0,\bg_T) = \rho_0(\bg_0)$.

While the formulation described above provides a static map $\Pi$, the Benamou-Brenier theorem \cite{benamou2000computational} shows that it is possible to find a dynamic transport map if transport costs are quadratic: $c(\bg_0,\bg_T) = ||\bg_0 - \bg_T||^2$. The dynamic transport map is specified by the time-dependent velocity field $\bm{v}^*(\bg,t)$, which transports the initial probability density $\rho(\bg_0,0) = \rho_0(\bg_0)$ according to the continuity equation $\pt \rho + \pmu (v_{\mu}^* \rho) = 0$. From a Lagrangian perspective, the deterministic state evolution of a particle is $dg_{\mu} = v_{\mu}^*(\bg,t)dt$. The theorem further shows that $\bm{v}^*$ can be written as the gradient of a time-dependent potential. 

The Schrödinger bridge problem (though originally posed as a large deviations problem) considers particles that evolve stochastically according to $dg_{\mu} = v_{\mu}(\bg,t)dt + \sqrt{\ve} dW_{\mu}$. The optimal velocity field $\bm{v}^*$ of the Schrödinger bridge problem can be found using stochastic control, where $\bm{v}^*$ is the optimal control policy that minimizes a quadratic control cost while guiding particles from the initial distribution $\rho_0$ to the final distribution $\rho_T$ \cite{chen2016relation}. We provide a derivation of the Schrödinger system of equations in Appendix \ref{sec:schrodinger}, which illustrates the stochastic control approach we take to derive our main result (Section \ref{sec:maintheorem}) in a simplified setting. The Schrödinger bridge problem is the entropy-regularized version of the Benamou-Brenier problem: in the limit $\ve \to 0$, the solution is identical to the one obtained from the Benamou-Brenier theorem. 

\section{Results}

\subsection{Growth-maximizing epigenetic landscapes} \label{sec:maintheorem}

We use a non-standard formulation of stochastic control \cite{chen2016relation,chen2021stochastic} (Appendix \ref{sec:schrodinger}) to find the growth-maximizing regulatory strategy that transports cell states from $\rho_0$ to $\rho_T$. In particular, we solve an \emph{inverse} problem where we find a terminal ``reward'' function $\omega(\bm{g})$ such that at optimality the normalized density at $T$ is precisely the target distribution $\rho_T(\bg)$ . If such a reward function and corresponding optimal strategy are found, the expected reward is $N_T^*\int \rho_T(\bg)\omega(\bm{g}) d\bm{g}$, where $N_T^*$ is the number of cells at time $T$. Since the optimal strategy maximizes expected reward, it is also the strategy that maximizes $N_T^*$ and thus the average growth rate $\Gamma^*_T = T^{-1} \log N_T^*$ amongst the class of strategies that guide a population from $\rho_0$ to $\rho_T$. 

We begin by defining $w(\bm{g},t)$ as the expected future reward of a cell with state $\bm{g}$ at time $t$. The expected reward at time $t$ can be written recursively in terms of the expected reward at time $t+dt$ by averaging over the cell's stochastic state evolution \eqref{eq:gmu} and growth \eqref{eq:gamma}, given the regulatory strategy $\bm{v}$. When the regulatory strategy maximizes expected reward, we have
\begin{align}
    w(\bm{g},t) = \max_{\bm{v}} \left\{ \left(1 + \gamma(\bm{g},\bm{v})dt \right)\langle w(\bm{g} + d\bm{g},t + dt) \rangle \right\},
\end{align}
where the expectation is over noise $dW_{\mu}$ and the growth term takes into account the probability that the cell doubles in interval $dt$. We expand $w(\bm{g} + d\bm{g},t + dt)$ in a Taylor series, compute expected values and retain terms of order $dt$ to get
\begin{align}
    \max_{\bm{v}} \left(\pt w + \left( f_{\mu} + v_\mu \right) \pmu w + \f2 \Sigma_{\mu \nu} \pmn w + \gamma w  \right) = 0, \label{eq:bellman}
\end{align}
where $\Sigma_{\mu \nu} = \sigma_{\mu \mu'}\sigma_{\nu \mu'}$. We have dropped the $\bm{g}, t$ arguments for convenience and a standard notation for partial derivatives is used ($\pmu = \partial/\partial g_{\mu}$, etc., unless specified otherwise). For $\gamma$ quadratic in $v_{\mu}$ \eqref{eq:gamma}, taking the max over $v_{\mu}$ we get $v_{\mu}^* = -h^{-1}_{\mu \nu} \partial_{\nu} \log w$. Plugging this back in \eqref{eq:bellman}, leads to the nonlinear PDE
\begin{align}
    \pt w + f_{\mu}\pmu w &- \frac{1}{2w} h^{-1}_{\mu \nu} (\pmu w) (\pnu w) \nonumber\\
    &+   \f2 \Sigma_{\mu \nu} \pmn w + \gamma_0 w  = 0. \label{eq:hjb}
\end{align}
The Cole-Hopf transform is used to transform nonlinear PDEs that arise in stochastic control to linear PDEs \cite{kappen2005linear, todorov2006linearly}. However, this transform is not suitable for linearizing \eqref{eq:hjb}. We instead consider a substitution $\psi = w^{1 + 1/\alpha}$, where $\alpha > 0$ is a constant. 

We find that this substitution yields a linear PDE in $\psi$ provided 
\begin{align}
    (h \Sigma)_{\mu \nu} = -\alpha \delta_{\mu \nu}.
\end{align}
Intuitively, this constraint requires that the magnitude of gene expression noise is inversely proportional to the cost of expressing the gene and that there is a single constant $\alpha$ that scales gene expression cost. Analogous constraints arise in path integral stochastic control \cite{kappen2005linear} and stochastic thermodynamics. The analogy to stochastic thermodynamics allows us to interpret $\alpha$ as an effective ``temperature'' parameter and the constraint $(h \Sigma)_{\mu \nu} = -\alpha \delta_{\mu \nu}$ as an analog of the Einstein relation in kinetic theory. The condition further implies $\Sigma_{\mu \nu}$ is full rank and $h$ is independent of $\bg$. We then have 
\begin{align}
\pt \psi + f_{\mu} \pmu \psi + \f2 \Sigma_{\mu \nu} \pmn \psi + \gamma_0\left(\frac{1+\alpha}{\alpha}\right)\psi  = 0. \label{eq:psi}
\end{align}
Equation \eqref{eq:psi} is the backward Kolmogorov equation \cite{gardiner1985handbook} of an uncontrolled process with drift $\bm{f}$, diffusion tensor $\Sigma/2$ and inhomogeneous growth rate $\gamma_0\left(\frac{1+\alpha}{\alpha}\right)$. Specifically, if the transition density of this uncontrolled process is $\phi(\bm{g},t;\bm{g}',t')$ for $t \ge t'$ (with boundary condition $\phi(\bm{g},t';\bm{g}',t') = \delta(\bm{g} - \bm{g}')$), we can express the solution of \eqref{eq:psi} in integral form:
\begin{align}
    \psi(\bm{g},t) = \int d \bg' \phi(\bg',T;\bg,t)\psi(\bg',T), \label{eq:psi_sol}
\end{align}
where $\psi(\bg,T) = \omega(\bg)^{1 + 1/\alpha}$ from our definition of $\psi$. Recall that under $\bm{v}^*$, $\rho$ satisfies the forward equation \eqref{eq:rho} with optimal control (expressed in $\psi$)
\begin{align}
    v_{\mu}^* = (1+\alpha)^{-1}\Sigma_{\mu \nu}\pnu \log \psi.  \label{eq:v_opt}
\end{align}
In the Schrödinger bridge problem (Appendix \ref{sec:schrodinger}), the density of the optimally controlled process $\rho$ can be expressed as the product of $\psi$ and another function $\hat{\psi}$ that satisfies the forward equation of the uncontrolled process. Motivated by this solution, we introduce $\hat{\psi}(\bg,t)$ defined as
\begin{align}
    \hat{\psi}(\bg,t) = \int d \bg' \phi(\bg,t;\bg',0)\hat{\psi}(\bg', 0).\label{eq:psihat_sol}
\end{align}
Given that $\phi$ is the transition density for the uncontrolled process introduced above, $\hat{\psi}$ satisfies the forward Kolmogorov equation \cite{gardiner1985handbook}
\begin{align}
    \pt \hat{\psi} + \pmu \left(f_{\mu} \hat{\psi}\right) - \f2 \Sigma_{\mu \nu} \pmn\hat{\psi}  - \gamma_0\left(\frac{1+\alpha}{\alpha}\right)\hat{\psi} = 0 \label{eq:psihat},
\end{align}
which is also the forward equation satisfied by $\phi$. We now ask if the (unnormalized) density $\rho(\bg,t)$ of the optimally controlled process can be expressed as a product of $\psi$ and $\hat{\psi}$. Unfortunately, a naive guess $\rho = \psi \hat{\psi}$ analogous to the solution of the Schrödinger bridge problem does not satisfy \eqref{eq:rho} and \eqref{eq:v_opt}. 

By considering various ansatz, we find that one particular ansatz
\begin{align}
    \rho(\bg,t) = \hat{\psi}(\bg,t)\psi(\bg,t)^{\frac{1}{1+\alpha}}, \label{eq:rho_relation}
\end{align}
does indeed satisfy \eqref{eq:rho} given \eqref{eq:v_opt}. Using \eqref{eq:psi}, \eqref{eq:psihat}, it is lengthy but straightforward to verify through direct substitution that \eqref{eq:rho_relation} satisfies \eqref{eq:rho}. 

In summary, the density $\rho$ of the optimally controlled process is given by \eqref{eq:rho_relation} in terms of $\psi$ and $\hat{\psi}$, which in turn satisfy \eqref{eq:psi}, \eqref{eq:psihat} respectively, and the boundary conditions
\begin{align}
 \rho_0(\bg) &= \hat{\psi}(\bg,0)\psi(\bg,0)^{\frac{1}{1+\alpha}}, \label{eq:bc1}\\
  N_T^*\rho_T(\bg) &= \hat{\psi}(\bg,T)\psi(\bg,T)^{\frac{1}{1+\alpha}}.  \label{eq:bc2}
\end{align}

$N_T^*$ is the number of cells at time $T$ under the growth-maximizing strategy $\bm{v}^*$ (given that we begin from one cell with state drawn from $\rho_0$ at $t=0$). Given a solution $\psi$ of the above set of equations, the optimal control is the gradient of a time-dependent potential \eqref{eq:v_opt}. However, it is unclear whether solutions exist for arbitrary boundary conditions \eqref{eq:bc1}, \eqref{eq:bc2}. In the special case $\gamma_0 = 0, \alpha \to 0, f_{\mu} = 0, \Sigma_{\mu \nu} = \ve \delta_{\mu \nu}$, the equations we obtain are identical to those for the Schrödinger bridge problem (Appendix \ref{sec:schrodinger}). Solutions are known to exist for arbitrary boundary conditions (under certain weak assumptions) and can be found using a simple procedure known as iterative proportional fitting or the Sinkhorn algorithm \cite{sinkhorn1964relationship,sinkhorn1967concerning,cuturi2013sinkhorn, peyre2019computational}. The Sinkhorm algorithm begins with an initial guess for $\hat{\psi}$ and iteratively updates \eqref{eq:bc1}, \eqref{eq:bc2} using \eqref{eq:psi_sol}, \eqref{eq:psihat_sol}. Whether an analogous procedure can be derived for our case is beyond the scope of this paper. 
\\

\begin{figure}[t]
    \center
    \includegraphics[scale=0.42]{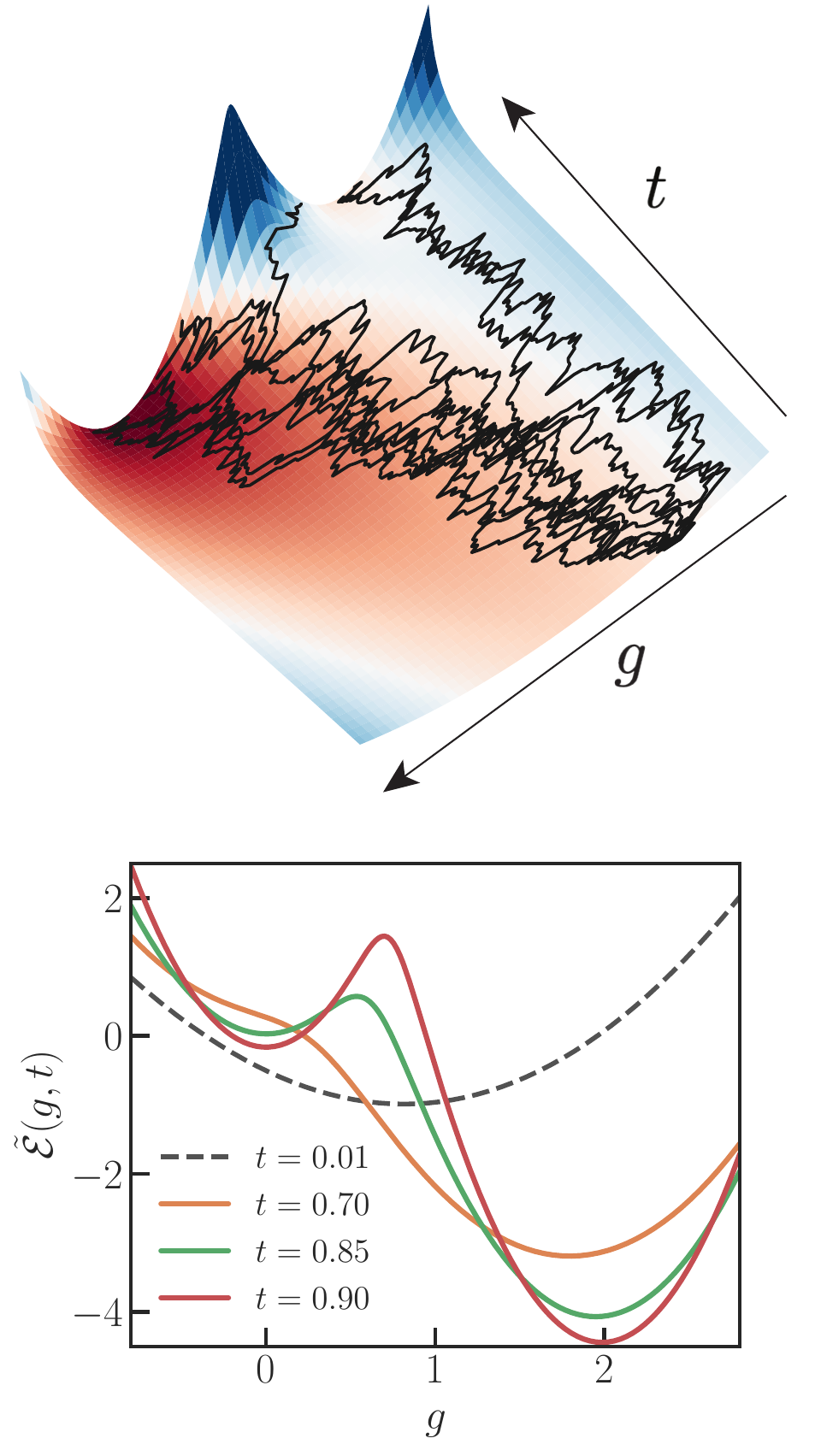}
    \caption{(Top) Sample trajectories on a one-dimensional time-varying potential landscape $\tilde{\mathcal{E}}$ differentiating into two cell types in a 1:4 ratio. Here $\tilde{\mathcal{E}} = \mathcal{E} + g^2/2 - t$ with $T = 1, \gamma_b =1, \lambda^2 = 0.1, \alpha = 0.1$. (Bottom) The landscape bifurcates from one stable state to two stable states at $t/T \approx 0.8$.}
    \label{fig:fig2}
\end{figure}

\subsection{Growth from a single cell} \label{sec:single}

An explicit expression for the potential landscape is obtained when the initial distribution is $\rho_0(\bg) = \delta(\bg - \bg_0)$, i.e., growth from a single cell with initial state $\bg_0$. In this case, $\hat{\psi}$ with
\begin{align}
\hat{\psi}(\bg,t) = \phi(\bg,t; \bg_0,0)/\psi(\bg_0,0)^{\frac{1}{1+\alpha}},
\end{align}
satisfies the initial condition \eqref{eq:bc1} (since $\phi(\bg,t'; \bg',t') = \delta(\bg - \bg')$) and the forward equation \eqref{eq:psihat}. From \eqref{eq:rho_relation}, the final distribution is
\begin{align}
\rho(\bg,T) = \phi(\bg,T; \bg_0,0)\left( \frac{\psi(\bg,T)}{\psi(\bg_0,0)} \right)^{\frac{1}{1+\alpha}}.
\end{align}
Enforcing $\rho(\bg,T) \propto \rho_T(\bg)$ simply requires choosing $\omega(\bg)$ such that
\begin{align}
    \psi(\bg,T) = \omega(\bg)^{\frac{1+\alpha}{\alpha}} =  \left(\frac{\rho_T(\bg)}{\phi(\bg,T;\bg_0,0)}\right)^{1+\alpha} \label{eq:rew_opt}.
\end{align}
From \eqref{eq:psi_sol} and \eqref{eq:v_opt}, the flow is then specified by a time-dependent potential $\mathcal{E} \equiv -(1+\alpha)^{-1} \log \psi$, i.e.,
\begin{align}
\mathcal{E}(\bg,t) = -\frac{1}{1+\alpha}\log \int d\bg' \phi(\bg',T;\bg,t)  \left(\frac{\rho_T(\bg')}{\phi(\bg',T;\bg_0,0)}\right)^{1+\alpha}.\label{eq:energy}
\end{align}
Equation \eqref{eq:energy} provides a method for computing $\mathcal{E}$ if we can either sample from or solve for the transition density $\phi(\bg',t';\bg,t)$ ($t' \ge t$). 
\\

\subsection{A 1D model of growth and differentiation} \label{sec:1d_main}

We illustrate our computational framework using a simplified one-dimensional model of growth and differentiation. We consider a single cell with initial state $g_0 = 0$ developing in time $T$ into a mixture of two Gaussians with means 0 and $\theta$. Cell state dynamics are given by 
\begin{align}
    dg = -gdt + v(g,t)dt + dW, \label{eq:1d_dyn}
\end{align}
with growth rate $\gamma(g,v) = \gamma_b - \f2 \lambda^2 g^2 - \f2 \alpha v^2$ for constants $\gamma_b,\lambda,\alpha$. This models the growth and diversification of a cell into a mixture of a fast-growing cell type ($g = 0$) and a slow-growing cell type ($g = \theta > 0$). The negative feedback term in \eqref{eq:1d_dyn} represents dilution and degradation of the marker gene $g$ and time is re-scaled by the timescale of dilution/degradation. 

\begin{figure}[t!]
    \center
    \includegraphics[scale=0.55]{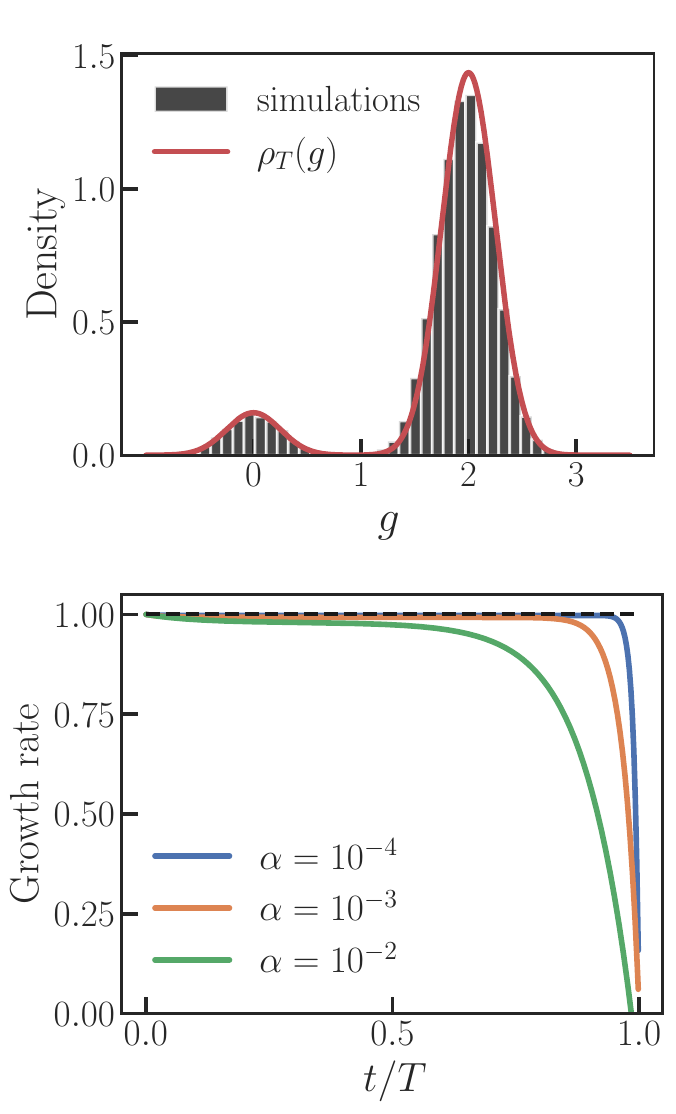}
    \caption{(Top) Samples from the 1D model match the target density. Here $\alpha = 0.1, \lambda^2 = 0.25, \gamma_b = 1$. (Bottom) Instantaneous population growth rates $\langle \gamma(g,v^*)\rangle_{g\sim \rho(.,t)} $ for different values of $\alpha$. For small values of $\alpha$, the growth and differentiation phases are clearly delineated. The black dashed curve shows $\gamma(0,0) = \gamma_b = 1$.}
    \label{fig:fig3}
\end{figure}

The transition density $\phi$ has an analytical expression (Appendix \ref{sec:1d_model}), and the landscape can be numerically evaluated for an arbitrary target density $\rho_T$ using \eqref{eq:energy}. Figure \ref{fig:fig2} shows sample trajectories on a representative potential landscape. The landscape is smooth and displays a saddle-node bifurcation from a single stable state to two stable states at late times (Figure \ref{fig:fig2}, bottom). Samples from this process match the target density (Figure \ref{fig:fig3}, top). The instantaneous growth rates for different $\alpha$ delineate two distinct phases of growth and differentiation for smaller $\alpha$ (Figure \ref{fig:fig3}, bottom). The optimal control optimizes for growth for most of the interval and differentiates close to the end by rapidly ramping up expression of $g$, which is cheaper in the long-run when $\alpha \ll 1$. 
\\

\subsection{A statistical bound on growth} \label{sec:bound}

A deep connection between optimal transport and stochastic thermodynamics allows for placing bounds on the entropy produced during a nonequilibrium process and for finding optimal protocols that saturate these bounds \cite{aurell2011optimal}. We now show that a similar bound can be derived in our formalism where growth rate plays a role analogous to entropy production. 

The number of cells at time $T$, $N_T^*$, or equivalently, the average growth rate $\Gamma^*_T = T^{-1} \log N_T^*$ (assuming we begin with one cell with state drawn from $\rho_0$) is self-consistently determined by the set of equations \eqref{eq:bc1}, \eqref{eq:bc2} for a given target $\rho_T$. A more interpretable form is obtained when $\Gamma^*_T$ is expressed in terms of the average growth rate $\Gamma_T^{\max}$ of a population (with initial density $\rho_0$) where the control is chosen to maximize growth over time $T$ with no constraint on the final density. The solution for this growth-maximizing process is obtained by setting $\omega(\bg) = \psi(\bg,T)^{\frac{\alpha}{1+\alpha}} = c > 0$, i.e., a constant positive reward $c$ is given for each cell at time $T$. $\Gamma_T^{\max}$ can be expressed in terms of
\begin{align}
    \bar{\nu}_T(\bg) &= \frac{1}{T}\int_0^T \nu(t;\bg)dt, \text{ where} \nonumber \\
    \quad \nu(t;\bg) &= \int d\bg' \gamma_0(\bg') \tilde{\phi}(\bg',t;\bg,0) 
\end{align}
is the instantaneous growth rate and $\tilde{\phi}$ is the normalized $\phi$ (Appendix \ref{sec:gamma_bound_deriv}). Specifically, we show that 
\begin{align}
\Gamma^{\max}_T = \frac{1}{T} \log \int d\bg \rho_0(\bg) e^{T\bar{\nu}_T(\bg)}. \label{eq:Nmax2}
\end{align} 
Plugging in $\rho_0(\bg) = \delta(\bg - \bg_0)$ in the above expression shows that $\bar{\nu}_T(\bg_0)$ is to be interpreted as the average growth rate of the growth-maximizing population that begins from a single cell with state $\bg_0$ and that $\phi(\bg,t;\bg_0,0)$ describes the time evolution of this process.

Multiplying both sides of \eqref{eq:bc1} by $e^{T\bar{\nu}_T(\bg)}$, integrating over $\bg$ and using \eqref{eq:Nmax2}, \eqref{eq:psi_sol}, \eqref{eq:bc2} (see Appendix \ref{sec:gamma_bound_deriv}), we get
\begin{align}
    \Gamma_T^* = \Gamma^{\max}_T - \frac{K_{\alpha}(\rho_T|| \rho^{\max}_T)}{T}, \label{eq:Gamma_bound}
\end{align}
where 
\begin{align}
    K_{\alpha}(\rho_T || \rho^{\max}_T)&\equiv \log \int d\bg e^{T\bar{\nu}_T(\bg)} \hat{\psi}(\bg,0)\xi(\bg)^{\frac{1}{1+\alpha}}, \label{eq:Kalph}\\
    \xi(\bg) &= \int d\bg' \phi(\bg',T;\bg, 0) \left(\frac{ \rho_T(\bg')}{\hat{\psi}(\bg',T)}\right)^{1+\alpha}, \nonumber
\end{align}
and $\rho^{\max}_T$ is the density at time $T$ if cells solely maximized growth beginning from $\rho_0$. Since the optimally controlled process specified by \eqref{eq:rho_relation} is the one that maximizes growth rate amongst the class of control strategies that transform the population of cells from $\rho_0$ to $\rho_T$, \eqref{eq:Gamma_bound} provides an upper bound on the average growth rate over all such control strategies. 

An application of Jensen's inequality shows that $K_{\alpha}(\rho_T || \rho^{\max}_T) \ge 0$ and equals zero only if $\rho_T = \rho^{\max}_T$ (Appendix \ref{sec:gamma_bound_deriv}). $K_{\alpha}$ thus appears to behave like a statistical divergence that measures the distance between the target density and the density at time $T$ if cells solely maximized growth. This interpretation is more transparent for the case when the developmental process begins from a single cell with initial state $\bg_0$ at $t = 0$ (Section \ref{sec:single}). The expression \eqref{eq:Kalph} for $K_{\alpha}$ in this case reduces to
\begin{align}
K_{\alpha}(\rho_T || \rho^{\max}_T) =  \frac{\alpha}{1+\alpha} D_{1+\alpha}(\rho_T || \rho^{\max}_T),
\end{align}
where $D_{1+\alpha}(\rho_T || \rho^{\max}_T)$ is the Rényi divergence \cite{van2014renyi} with parameter $1+\alpha$ and $\rho^{\max}_T(\bg) = \tilde{\phi}(\bg,T;\bg_0,0)$. The well-known Kullback-Leibler divergence is recovered in the limit $\alpha \to 0$. 
\\

\section{Conclusion} 

A longstanding challenge in biological physics is to build interpretable, quantitative models of complex cellular regulatory processes that reflect general biological principles. Recent work has made progress towards this goal by describing experimental data using the language of dynamical systems theory, which emphasizes global geometric features at the expense of molecular details. An underlying assumption is that gene regulatory dynamics can be described using relatively simple gradient-like dynamics. Here, we show that regulatory circuits that maximize growth during cellular diversification indeed correspond to gradient flow on a potential landscape under certain assumptions on how gene regulation impacts growth. In deriving this result, we have ignored various important aspects of development, including intercellular communication and spatial structure. A general framework that incorporates these features remains elusive, though analogies between development processes and high-dimensional generative modeling offer some hope that this is feasible. 

Our theory has close connections with optimal transport, and the Benamou-Brenier formalism in particular. The quadratic control transport cost considered in the Benamou-Brenier formalism directly relates to entropy produced during a nonequilibrium process, establishing a connection between optimal transport and stochastic thermodynamics \cite{aurell2011optimal,aurell2012refined,van2023thermodynamic}. Like the Benamou-Brenier formalism, we identify a set of constraints under which finding a transport map reduces to finding an appropriate time-dependent potential. Our results appear identical in the special case of $\gamma_0(\bg) = 0, \alpha \to 0$, though further analysis is required to clarify the relationship between these two formalisms and unbalanced variants of optimal transport \cite{chizat2018unbalanced}. This connection also opens up the possibility of fitting single-cell gene expression data using growth-informed transport costs that naturally incorporate growth and offer a clearer biological interpretation. 

\begin{acknowledgments}
G.R. thanks Wenping Cui, Emmy Blumenthal and Pankaj Mehta for valuable comments on the manuscript. This work was partially supported by a joint research agreement between Princeton University and NTT Research Inc. 
\end{acknowledgments}

\appendix

\section{The Schrödinger bridge problem and the Benamou-Brenier theorem}\label{sec:schrodinger}
We provide a brief review of the Schrödinger bridge problem. We follow ref. \cite{chen2021stochastic, chen2016relation} and derive the solution by formulating the bridge problem as a stochastic control problem. We use this example as a simpler version of our main result, and to highlight points where our derivation differs from that of the Schrödinger bridge setting. 

We follow the notation used in the main text. Consider the state evolution equation
\begin{align}
    dg_{\mu} = v_{\mu}(\bg, t)dt + \sqrt{\ve}dW_{\mu}.
\end{align}
The density $\rho(\bg,t)$ satisfies the forward equation
\begin{align}
    \partial_t \rho = -\pmu (v_{\mu}\rho) + \frac{\ve}{2}\pmu \pmu  \rho, \label{eq:schro_rho}
\end{align}
where $\pmu \pmu$ is the Laplacian. Here, we have assumed a simpler form for the uncontrolled dynamics ($v_{\mu} = 0$) as a diffusion with independent and identical noise contributions. 

In the Schrödinger bridge problem, we aim to find the optimal control $\bm{v}^*$ that guides a particle drawn from initial distribution $\rho_0(\bg)$ to a final distribution $\rho_T(\bg)$ in finite time $T$. In contrast to our growth maximization objective, we minimize the quadratic control cost: $\langle \frac{1}{2}\int_0^T ||\bm{v}(.,t)||^2 dt \rangle$, such that the distributions at $t = 0$ and $t = T$ are $\rho_0$ and $\rho_T$ respectively. The expectation is taken over the particle's trajectory. It can be shown that minimizing the quadratic control cost is equivalent to minimizing the Kullback-Leibler (KL) divergence between the controlled process and the uncontrolled Brownian motion, subject to the boundary constraints. In other words, the Schrödinger bridge problem aims to find the process that is closest (in the KL-divergence sense) to a reference process (here, Brownian motion) that still satisfies the density constraints at $t =0$ and $t= T$. 

We find the optimal control $\bm{v}^*$ by minimizing the quadratic control cost while maximizing the terminal reward $\omega(\bg)$. The hitherto unknown $\omega(\bg)$ is to be chosen such that $\rho(\bg,T) = \rho_T(\bg)$ under $\bm{v}^*$. We define  $z(\bg, t)$ as the expected reward given that the particle's state is $\bg$ at time $t$. Writing down the Bellman equation for $z$, we get
\begin{align}
    z(\bg,t) = \max_{\bm{v}} \left\{-\frac{1}{2}v_{\mu} v_{\mu} dt + \langle z(\bg + d\bg, t+dt) \rangle \right\}. \label{eq:z}
\end{align}
Expanding to second order and taking the expectation, $\langle z(\bg + d\bg, t+dt) \rangle = dt(\pt z + v_{\mu} \pmu z + \frac{\ve}{2}\pmu \pmu z )$. Plugging this back into \eqref{eq:z} and taking the max over $\bm{v}$, the optimal control is
\begin{align}
    v^*_{\mu}(\bg,t) = \pmu z(\bg,t)
\end{align}
Plugging the above back into \eqref{eq:z} and simplifying, we have
\begin{align}
    \pt z + \frac{1}{2}(\pmu z)(\pmu z) + \frac{\ve}{2} \pmu \pmu z = 0,
\end{align}
with boundary condition $z(\bg,T) = \omega(\bg)$. Substituting $z(\bg,t) = \ve \log \chi(\bg,t)$, we get a linear PDE in $\chi$:
\begin{align}
    \pt \chi + \frac{\ve}{2} \pmu \pmu \chi = 0. \label{eq:chi}
\end{align}
Define $\hat{\chi}$ as the solution to the PDE
\begin{align}
    \pt \hat{\chi} - \frac{\ve}{2} \pmu \pmu \hat{\chi} = 0. \label{eq:chihat}
\end{align}
The boundary conditions $\chi(\bg,T)$ and $\hat{\chi}(\bg,0)$ are to be found (which also fixes $\omega(\bg)$). Direct substitution shows that 
\begin{align}
    \rho(\bg,t) = \chi(\bg,t)\hat{\chi}(\bg,t)
\end{align}
together with
\begin{align}
    v^*_{\mu}(\bg,t) = \ve \pmu  \log \chi(\bg,t)
\end{align}
satisfies \eqref{eq:schro_rho}. Thus, the Schrödinger bridge problem can be solved by finding boundary conditions $\chi(.,T)$ and $\hat{\chi}(.,0)$ such that
\begin{align}
    \rho_0(\bg) &= \chi(\bg,0)\hat{\chi}(\bg,0),\\
    \rho_T(\bg) &= \chi(\bg,T)\hat{\chi}(\bg,T), 
\end{align}
where $\chi(.,0)$ is obtained from $\chi(.,T)$ using the backward equation \eqref{eq:chi} and $\hat{\chi}(.,T)$ is obtained from $\hat{\chi}(.,0)$ using the forward equation \eqref{eq:chihat}. 

The Benamou-Brenier theorem is obtained in the limit $\ve \to 0$. Note that one can obtain a \emph{deterministic} flow $\bm{u}^*(\bg,t)$ by substituting
\begin{align}
    u_{\mu}^* = v_{\mu}^* - \frac{\ve}{2} \pmu \log \rho
\end{align}
into \eqref{eq:schro_rho}. The continuity equation under the deterministic optimal control then reads 
\begin{align}
    \partial_t \rho + \pmu (u_{\mu}^*\rho) = 0.
\end{align}

\section{A one-dimensional model of growth and differentiation} \label{sec:1d_model}

We consider a one-dimensional case where $\gamma(g,v) = \gamma_b - \f2 \lambda^2 g^2 - \f2 \alpha v^2$ and $f(g,t) = -g$. The target density is a mixture of two Gaussians: $\rho_T(g) = p\varphi(g/\kappa)/\kappa + (1-p) \varphi((g-\theta)/\kappa)/\kappa$ with $\theta > 0$, where $\varphi$ is the standard normal density and $\kappa$ is the standard deviation.  This corresponds to the case when a cell that differentiates into a cell type with $|g| > 0$ incurs a growth cost quantified by $\lambda^2$. Here, we derive the transition density $\phi(g,t;g',t')$ for the uncontrolled process.

From \eqref{eq:psihat}, $\phi(g,t;g',t')$ satisfies
\begin{align}
    \pt \phi = \pg \left( g\phi \right) + \f2 \pgg \phi + \left(\gamma_b - \f2 \lambda^2 g^2\right)\left(1 + \frac{1}{\alpha} \right) \phi. 
\end{align}
Define $\gamma_b' = \gamma_b(1 + 1/\alpha)$ and $\lambda'^2 = \lambda^2(1 + 1/\alpha)$. Substituting $\phi(g,t;g',t') = \bar{\phi}(g,t;g',t') e^{\gamma_b' t}$ gives 
\begin{align}
    \pt \bar{\phi} = \pg \left( g\bar{\phi} \right) + \f2 \pgg \bar{\phi} - \f2 \lambda'^2 g^2\bar{\phi}. 
\end{align}
Substituting 
\begin{align}
    \bar{\phi}(g,t;g',t') = \hat{\phi}(g,t;g',t')e^{(\sqrt{1 + \lambda'^2} - 1)(g^2 - t)/2}
\end{align} gives
\begin{align}
    \pt \hat{\phi} = \pg \left(\sqrt{1+ \lambda'^2} g\hat{\phi} \right) + \f2 \pgg \hat{\phi},
\end{align}
which is the forward equation of an Ornstein-Uhlenbeck process with drift $-\sqrt{1+\lambda'^2}g$ and diffusion coefficient $1/2$. The solution is well-known
\begin{align}
    \hat{\phi}(g,t;g',t') &= \frac{(1+\lambda'^2)^{1/4}}{\sqrt{\pi \left( 1 - e^{-2\sqrt{1+\lambda'^2}(t-t')}\right)}} \nonumber \\
    &\times e^{- \frac{\sqrt{1+\lambda'^2}\left(g - g'e^{-\sqrt{1+\lambda'^2}(t-t')}\right)^2}{ 1 - e^{-2\sqrt{1+\lambda'^2}(t-t')}}}. 
\end{align}
The prefactor is set by the boundary condition $\phi(g,t';g',t') = \delta(g-g')$. We get
\begin{align}
    \phi(g,t;g',t') &= e^{\left(\gamma_b' - \frac{\sqrt{1+\lambda'^2} -1}{2}\right)(t-t') +\left(\frac{\sqrt{1+\lambda'^2} -1}{2}\right)(g^2-g'^2) }\nonumber\\
    &\times \hat{\phi}(g,t;g',t'). 
\end{align}

\section{Derivation of eq. \eqref{eq:Gamma_bound}}\label{sec:gamma_bound_deriv}

We first consider a process where the control variable $\bm{v}$ is chosen to maximize the average growth rate over time $T$ with initial density $\rho_0(\bg)$ and with no constraint on the final density. We will use a prime symbol to distinguish this growth-maximizing process from our original one. The solution to this process is obtained by setting $\omega'(\bg) = 1$ (or any positive constant). Note that it is only the boundary conditions for $\psi'$ and $\hat{\psi}'$ that change and not their equations \eqref{eq:psi}, \eqref{eq:psihat}. Also note that $\phi$ remains the same. 

Since $\psi' = w'^{1 + 1/\alpha}$, $\omega(\bg) = 1$ corresponds to the boundary condition $\psi'(\bg,T) = 1$ for all $\bg$. Let
\begin{align}
\bar{N}_t(\bg) \equiv \int d\bg' \phi(\bg',t; \bg,0).\label{eq:phi_num_app}
\end{align}
From \eqref{eq:psi_sol}, since $\psi'(\bg,T) = 1$, we have $\psi'(\bg,t) = \bar{N}_{T-t}(\bg)$ and in particular $\psi'(\bg,0) = \bar{N}_{T}(\bg)$. From \eqref{eq:bc2}, the density at $T$ is $\hat{\psi}'(\bg,T)$, which from \eqref{eq:psihat_sol} is given by 
\begin{align}
\hat{\psi}'(\bg,T) &= \int d\bg'  \phi(\bg, T; \bg',0) \hat{\psi}'(\bg',0), \nonumber \\
&= \int d\bg' \phi(\bg, T; \bg',0) \left( \frac{\rho_0(\bg')}{\bar{N}_T(\bg')^{\frac{1}{1+\alpha}}} \right) \label{eq:psihatprime_app}
\end{align}
The number of cells at time $T$ for the growth-maximizing process is
\begin{align}
N_T^{\max} &= \int d\bg \rho(\bg,T) \nonumber \\
&= \int d\bg \hat{\psi}'(\bg,T) \nonumber\\
&= \int d\bg' \rho_0(\bg') \bar{N}_T(\bg')^{\frac{\alpha}{1+\alpha}},\label{eq:Nmax_app}
\end{align} 
where we have used \eqref{eq:bc2} in the first step and \eqref{eq:psihatprime_app}, \eqref{eq:phi_num_app} in the second step when integrating over $\bg$. 

We can express $\bar{N}_t(\bg)$ in terms of a growth rate. Since $\phi$ satisfies \eqref{eq:psihat}, integrating both sides of \eqref{eq:psihat} over $\bg$ and setting boundary terms at $\pm \infty$ to zero, we have
\begin{align}
\frac{ d \bar{N}_t(\bg)}{dt} &= \left( \frac{1+\alpha}{\alpha} \right) \int d\bg' \gamma_0(\bg') \phi(\bg',t; \bg,0)\nonumber\\
&=  \left( \frac{1+\alpha}{\alpha} \right)  \bar{N}_t(\bg) \int d\bg' \gamma_0(\bg') \tilde{\phi}(\bg',t; \bg,0),\nonumber \\
&=  \left( \frac{1+\alpha}{\alpha} \right)  \bar{N}_t(\bg) \nu(t;\bg) \label{eq:N_diff_app}
\end{align}
where we have defined the normalized density $\tilde{\phi}$ and $\nu(t;\bg)$ is the population averaged growth rate for the uncontrolled process given that the population began at $t = 0$ from a single cell with state $\bg$:
\begin{align}
 \nu(t;\bg) \equiv \int d\bg' \gamma_0(\bg') \tilde{\phi}(\bg',t; \bg,0)
\end{align}
Integrating \eqref{eq:N_diff_app}, we get
\begin{align}
\frac{1}{T}\log \bar{N}_T(\bg) &= \left( \frac{1+\alpha}{\alpha} \right) \left(  \frac{1}{T} \int_0^T \nu(t;\bg) dt \right) \nonumber\\
&\equiv \left( \frac{1+\alpha}{\alpha} \right)\bar{\nu}_T(\bg). \label{eq:Nbar_app}
\end{align}
Plugging the above expression for $\bar{N}_T(\bg)$ into \eqref{eq:Nmax_app}, we get 
\begin{align}
\Gamma^{\max}_T = \frac{1}{T} \log N_T^{\max} = \frac{1}{T} \log \int d\bg \rho_0(\bg) e^{T\bar{\nu}_T(\bg)}. \label{eq:Nmax2_app}
\end{align} 
\\

We now express the average growth rate $\Gamma^*_T = T^{-1} \log N_T^*$ of the optimally controlled process in terms of $\Gamma^{\max}_T$ and the divergence $K_{\alpha}$ defined in the main text. We return to the original process in $\psi$ and $\hat{\psi}$. Plugging in $\rho_0(\bg) = \hat{\psi}(\bg,0)\psi(\bg,0)^{\frac{1}{1+\alpha}}$ into the right hand side of \eqref{eq:Nmax2_app}, we have
\begin{align}
&\int d\bg \rho_0(\bg) e^{T\bar{\nu}_T(\bg)} = \int d\bg e^{T\bar{\nu}_T(\bg)} \hat{\psi}(\bg,0)\psi(\bg,0)^{\frac{1}{1+\alpha}} ,\\
&= \int d\bg e^{T\bar{\nu}_T(\bg)} \hat{\psi}(\bg,0)\left( \int d\bg' \phi(\bg',T;\bg, 0) \left(\frac{N_T^* \rho_T(\bg')}{\hat{\psi}(\bg',T)}\right)^{1+\alpha}\right)^{\frac{1}{1+\alpha}}\\
&= N_T^*  \int d\bg e^{T\bar{\nu}_T(\bg)} \hat{\psi}(\bg,0)\xi(\bg)^{\frac{1}{1+\alpha}},
\end{align}
where we have used \eqref{eq:psi_sol}, \eqref{eq:bc2} in the second step and defined
\begin{align}
    \xi(\bg) \equiv \int d\bg' \phi(\bg',T;\bg, 0) \left(\frac{ \rho_T(\bg')}{\hat{\psi}(\bg',T)}\right)^{1+\alpha}.
\end{align}
Re-arranging and expressing in terms of growth rates, 
\begin{align}
\Gamma_T = \Gamma_T^{\max} - \frac{K_{\alpha}(\rho|| \rho^{\max})}{T}\label{eq:Gamma_app}
\end{align}
where 
\begin{align}
K_{\alpha}(\rho_T || \rho^{\max}) \equiv \log \int d\bg e^{T\bar{\nu}_T(\bg)} \hat{\psi}(\bg,0)\xi(\bg)^{\frac{1}{1+\alpha}}. \label{eq:Kalph_app}
\end{align}
To show $K_{\alpha}(\rho_T || \rho^{\max}) \ge 0$, we rewrite the above expression in terms of normalized densities. We have $\phi(\bg',T;\bg,0) = \bar{N}_T(\bg) \tilde{\phi}(\bg',T;\bg,0)$ and express $\hat{\psi}$ in terms of its normalized density, $\hat{\psi}(\bg',T) = \hat{N}_T \hat{\psi}_n(\bg',T)$, where the normalization factor is
\begin{align}
\hat{N}_T &= \int d \bg' \hat{\psi}(\bg',T) \nonumber \\
&= \int d\bg' \int d\bg \phi(\bg',T;\bg,0)  \hat{\psi}(\bg,0) \nonumber\\
&=   \int d \bg \bar{N}_T(\bg) \hat{\psi}(\bg,0).
\end{align}
Using $\log \bar{N}_T(\bg) = \left(\frac{1+\alpha}{\alpha}\right) T\bar{\nu}_T(\bg)$, we have
\begin{align}
K_{\alpha}(\rho_T || \rho^{\max}) &= \log \int d\bg \bar{N}_T(\bg)^{\frac{\alpha}{1+\alpha}} \hat{\psi}(\bg,0)\xi(\bg)^{\frac{1}{1+\alpha}} \nonumber\\
&= \log \int d\bg \frac{\bar{N}_T(\bg) \hat{\psi}(\bg,0)}{\int d \bg'' \bar{N}_T(\bg'') \hat{\psi}(\bg'',0) }\tilde{\xi}(\bg)^{\frac{1}{1+\alpha}},\nonumber \\ 
&= \log \int d\bg Q(\bg) \tilde{\xi}(\bg)^{\frac{1}{1+\alpha}},\label{eq:Kalph_main_app}
\end{align}
where we have defined 
\begin{align}
Q(\bg) &\equiv \frac{\bar{N}_T(\bg) \hat{\psi}(\bg,0)}{\int d \bg' \bar{N}_T(\bg') \hat{\psi}(\bg',0) }, \\
\tilde{\xi}(\bg) &\equiv \int d\bg' \tilde{\phi}(\bg',T;\bg, 0) \left(\frac{ \rho_T(\bg')}{\hat{\psi}_n(\bg',T)}\right)^{1+\alpha}. 
\end{align}
Using Jensen's inequality, 
\begin{align}
 &\left(\int d\bg' \tilde{\phi}(\bg',T;\bg, 0) \left(\frac{ \rho_T(\bg')}{\hat{\psi}_n(\bg',T)}\right)^{1+\alpha}\right)^{\frac{1}{1+\alpha}} \nonumber\\
 &\ge  \int d\bg' \tilde{\phi}(\bg',T;\bg, 0)\frac{ \rho_T(\bg')}{\hat{\psi}_n(\bg',T)}, \label{eq:Kalph_ineq_app}
\end{align}
with equality only if $\rho_T(\bg) = \hat{\psi}_n(\bg,T) \propto \hat{\psi}(\bg,T)$ for all $\bg$. Note that from \eqref{eq:bc2} the equality condition $$\rho_T(\bg) \propto \hat{\psi}(\bg,T)$$ implies $\psi(\bg,T)$ (and thus $\omega(\bg)$) is a constant, which corresponds precisely to the process that maximizes growth rate. Plugging \eqref{eq:Kalph_ineq_app} back in  \eqref{eq:Kalph_main_app} and using the definition of $Q(\bg)$, we get $K_{\alpha}(\rho_T || \rho^{\max}) \ge 0$.

\bibliography{main}% Produces the bibliography via BibTeX.

\end{document}